\documentclass[runningheads]{llncs}
\usepackage[T1]{fontenc}
\usepackage{booktabs}
\usepackage{hyperref}       %
\usepackage{graphicx}
\usepackage{color}

\usepackage{multirow}
\usepackage{multicol}
\usepackage[table]{xcolor}
\usepackage{algorithm}
\usepackage{algpseudocode}

\algrenewcommand\algorithmicrequire{\textbf{Input:}}
\algrenewcommand\algorithmicensure{\textbf{Output:}}

\usepackage{todonotes}
\usepackage{amsmath}

\usepackage{amstext}
\usepackage{amsmath}
\usepackage{enumitem}
\usepackage{dblfloatfix}
\usepackage{wrapfig}
\usepackage{tikz}
\usepackage{pgfplots}
\usepackage{subcaption}
\usepackage{pgfplotstable}
\usetikzlibrary{patterns}

\renewcommand{\bf}{\fontseries{b}\selectfont}

\newcommand{\sys}[1]{\textsc{Gar}\def\temp{#1}\ifx\temp\empty{}\else\raisebox{-.4ex}{\scriptsize #1}\fi}
\newcommand{\slidesys}[1]{\textsc{SlideGar}\def\temp{#1}\ifx\temp\empty{}\else\raisebox{-.4ex}{\scriptsize #1}\fi}
\newcommand{\sysbm}[1]{\sys{}${}_{BM25}$}
\newcommand{\gar}{\textsc{Gar}}
\newcommand{\sgar}{\textsc{SlideGar}}
\newcommand{\mpara}[1]{\medskip\noindent{\textbf{#1}}}
\newcommand{\quam}{\textsc{Quam}}

\begin{document}
\title{Guiding Retrieval using LLM-based Listwise Rankers}

\author{Mandeep Rathee\inst{1}\orcidID{0000-0002-7339-8457} \and
Sean MacAvaney\inst{2}\orcidID{0000-0002-8914-2659} \and
Avishek Anand\inst{3}\orcidID{0000-0002-0163-0739} }

\institute{L3S Research Center, Hannover, Germany \\
\email{rathee@l3s.de} \and
University of Glasgow, Glasgow, UK \\
\email{ sean.macavaney@glasgow.ac.uk} \and
Delft University of Technology, Delft, The Netherlands\\
\email{avishek.anand@tudelft.nl}
}

\maketitle              %
\begin{abstract}

Large Language Models (LLMs) have shown strong promise as rerankers, especially in ``listwise'' settings where an LLM is prompted to rerank several search results at once. However, this ``cascading'' retrieve-and-rerank approach is limited by the bounded recall problem: relevant documents not retrieved initially are permanently excluded from the final ranking.
Adaptive retrieval techniques address this problem, but do not work with listwise rerankers because they assume a document's score is computed independently from other documents.
In this paper, we propose an adaptation of an existing adaptive retrieval method that supports the listwise setting and helps guide the retrieval process itself (thereby overcoming the bounded recall problem for LLM rerankers).
Specifically, our proposed algorithm merges results both from the initial ranking and feedback documents provided by the most relevant documents seen up to that point. Through extensive experiments across diverse LLM rerankers, first stage retrievers, and feedback sources, we demonstrate that our method can improve nDCG@10 by up to 13.23\% and recall by 28.02\%--all while keeping the total number of LLM inferences constant and overheads due to the adaptive process minimal. The work opens the door to leveraging LLM-based search in settings where the initial pool of results is limited, e.g., by legacy systems, or by the cost of deploying a semantic first-stage.

\end{abstract}

\keywords{Reranking  \and Adaptive Retrieval \and LLM}

\section{Introduction}

One of the most prevalent approaches in document search systems is the two-stage retrieve-and-rerank (``cascading'') pipeline, where an initial retrieval phase is followed by a more involved reranker~\cite{matveeva2006high}. 
In this paradigm, the results from a retrieval model are re-ordered by a reranker---a slower but more capable relevance model---based on their estimated relevance to the query. 
For example, LLM-based rankers have recently gained prominence due to their ability to understand and process complex language patterns, often performing effectively in listwise reranking settings where the goal is to produce an ordered list of documents without explicit relevance scores for individual documents~\cite{sun2023chatgpt,ma2023zero,pradeep2023rankvicuna,pradeep2023rankzephyr}.
However, a significant limitation of reranking systems is the \textit{bounded recall problem}. Specifically, if a relevant document fails to be retrieved in the initial retrieval phase, it is irrevocably excluded from the final ranked list, regardless of its potential relevance. This constraint limits the overall effectiveness of cascading systems, as the presence of relevant documents in the retrieved set is a prerequisite for successful reranking.

Recently, Adaptive Retrieval (AR) has emerged as a promising solution to mitigate the bounded recall problem~\cite{macavaney2022adaptive}. Unlike conventional reranking, which solely reorders the initially retrieved documents, AR dynamically expands the retrieval set by leveraging the reranker's assessments during the reranking process itself. 
Instead of reranking all retrieved documents indiscriminately, AR applies the Clustering Hypothesis, which suggests that similar documents tend to be relevant to the same query~\cite{jardine1971use}, to selectively identify unretrieved documents that are similar to those identified as highly relevant by the reranker. AR is a form of Pseudo-Relevance Feedback (PRF), though unlike traditional PRF that applies query rewriting, the query remains unchanged and the index is not re-queried. AR facilitates this property by perfomring lookups in a corpus graph that encodes the top-k similarity relationships between documents. By doing so, AR enhances the likelihood of including relevant documents that may have been omitted in the initial retrieval, thereby improving the overall recall of the system.
By constructing the corpus graph offline, overheads of applying AR (as compared to traditional reranking) are minimal, since lookups can be conducted in a constant and minimal time.\footnote{In contrast, traditional PRF methods, such as as RM3~\cite{abdul2004umass}, rewrite and re-execute the query. This is both a costly process and one that can lead to query drift.}

Previous studies have demonstrated that AR consistently enhances recall across a variety of retrieval and reranking models~\cite{macavaney2022adaptive,kulkarni2023lexically,macavaney2024reproducibility}, underscoring its potential as a robust enhancement to traditional ranking pipelines. However, existing work in AR made the Probably Ranking Principle (PRP~\cite{robertson1977probability}) assumption that a relevance score is a function solely of the query and document. LLM reranking models use signals from an entire ranked list to determine relevance, breaking the PRP assumption. Therefore, in this work, we build on the foundations of AR to explore their applicability to listwise Large Language Model (LLM) rerankers.

The primary challenge in integrating AR with LLM rerankers lies in the fundamental difference between pointwise and listwise ranking approaches. 
So far, AR has been designed for pointwise ranking methods that score each document independently based on its relevance to the query like BERT, or MonoT5~\cite{nogueira2020document,khattab2020colbert}. 
In contrast, LLM rerankers excel in listwise settings, where the ranking function considers the relative order of documents as a whole rather than assigning individual scores. This discrepancy necessitates the adaptation of AR to accommodate the listwise nature of LLM rerankers.

In this paper, we make several key contributions to address this challenge:

\begin{itemize}

\item We propose \sgar{} to modify AR for compatibility with ranking functions that output only a ranked order of documents, rather than discrete relevance scores. 

\item We conduct extensive experiments across a diverse array of LLM rankers and substrates to evaluate the effectiveness and efficiency of AR in enhancing LLM-based reranking performance. Our experimental results show that \sgar{} can improve the nDCG@10 by up to $13.23\%$ and recall by $28.02\%$ over the state-of-the-art listwise rankers.

\item To further understand the computational overheads, we perform ablation studies and find that \sgar{} adds negligible latency cost (only 0.02\% of rankers latency) in the re-ranking pipelines. 

\end{itemize}

Our findings reveal that integrating AR with LLM rankers improves recall and nDCG scores across most corpus graphs and LLM configurations. These results affirm the viability of AR as a valuable enhancement to LLM-based reranking systems, offering a pathway to more effective document ranking, even in settings where the first stage is limited (e.g., due to legacy systems or due to the cost of deploying a semantic first stage.). We release our code at \url{https://github.com/Mandeep-Rathee/llmgar}.

\section{Related Work}

\subsection{Retrieval and Ranking}

Cascading approaches have a long history in information access systems~\cite{matveeva2006high}. 
In retrieval systems, cascading often manifests itself as a retrieve-then-rerank pipeline. 
Sparse retrieval algorithms (such as BlockMax-WAND~\cite{ding2011faster}) and dense retrieval algorithms (such as HNSW~\cite{malkov2018efficient}) can produce an initial set of (potentially approximate) top-$k$ search results in sublinear time with respect to the size of the corpus, allowing them to scale to massive corpora. These initial retrieval results are limited by the representation bottleneck (e.g., the terms matched in a bag-of-words representation), and therefore often benefit from an additional iteration of refinement before being presented to a user---often through a cross-encoding neural model, such as one based on BERT~\cite{nogueira2020document} or an LLM~\cite{sun2023chatgpt}. By considering the content of the query and document jointly, they can often produce a higher-quality relevance ranking. 
These models are not capable of scaling to large corpora, given that they operate in linear (or superlinear) time (usually with high constant factors).
The central limitation of cascading approaches is the bounded recall problem; documents filtered out in previous stages have no chance of being seen by the reranker. In the context of search systems, this has spawned substantial research into stronger first-stage retrieval systems, such as those that leverage dense~\cite{lin2021batch} or learned sparse vectors~\cite{nguyen2023unified}. Despite considerable progress in this space, these methods still struggle with certain queries~\cite{weller2024followir} due to their representation bottlenecks. Furthermore, such systems often depend on an entirely new core engine, which can be operationally challenging when existing infrastructure has been built around a lexical-based engine.

\subsection{Adaptive Retrieval}\label{sec:back-ada}

To overcome the recall limitations of cascading approaches, adaptive reranking~\cite{macavaney2022adaptive} techniques have been proposed. These methods utilize an index structure based on document-document similarities and leverage relevance estimations obtained during the reranking process to retrieve and score documents that were not captured in the initial retrieval stage. 
Drawing inspiration from the Cluster Hypothesis~\cite{jardine1971use}, adaptive reranking posits that documents in close proximity to those with high relevance scores are also likely to be relevant; by pulling in these similar documents when reranking, recall limitations of the first stage can be mitigated. By using a document-to-document similarity graph (``corpus graph'', akin to HNSW~\cite{malkov2018efficient}), this process can be accomplished with minimal overheads.

Several strategies for adaptive reranking have been proposed. For example, \gar{}~\cite{macavaney2022adaptive} and \quam{}~\cite{rathee2024quam} employ an alternating strategy that scores batches of documents from both the initial retrieval pool and those sourced from the corpus graph.
Beyond cross-encoders, adaptive reranking techniques have also been successfully applied to bi-encoders~\cite{kulkarni2023lexically,macavaney2024reproducibility} and ensemble models~\cite{yang2024cluster}. For instance, LADR~\cite{kulkarni2023lexically} uses an efficient lexical model to identify promising ``seed'' documents, which are then used to explore the corpus graph further.
AR is inherently designed for pointwise ranking methods. In this work we extend AR to listwise settings, where the ranking function considers the relative order of documents as a whole rather than assigning individual scores. 

\subsection{Pseudo-Relevance Feedback (PRF)}
Another line of work, pseudo-relevance feedback (PRF), tries to improve recall by providing feedback from the top-$k$ initial retrieved documents. Examples of methods based on this approach include Rocchio~\cite{rocchio1971relevance}, KL expansion~\cite{zhai2001model}, relevance modeling~\cite{metzler2005markov}, LCE~\cite{metzler2007latent}, and RM3 expansion~\cite{abdul2004umass}. These methods result in the \textit{query drift} problem. Recent works~\cite{mackie2023generative,mackie2023generativeall} have proposed Generative Relevance Feedback (GRF). The goal is to enhance the original query by using an LLM. The expanded query can incorporate the original terms along with useful terms generated from various subtasks, such as keywords, entities, chain-of-thought reasoning, facts, news articles, documents, and essays. However, such methods of expanding LLM-based queries are costly, since the number of output tokens is high.

\subsection{LLM Rerankers}
Recent works, such as~\cite{sun2023chatgpt,sun2024investigation}, have proposed \textbf{RankGPT}, zero-shot LLMs for document ranking task. The zero-shot LLM-based ranker shows significant improvements in ranking tasks over the various benchmark datasets. The LLM-based ranker is instructed by providing a prompt and takes the query and the text of the documents as input (based on the size of the context) and ranks these documents. Recently, Large Language Models (LLMs) have been shown to be highly effective in the reranking phase. \textbf{RankLLaMA}~\cite{ma2024fine} leverages the LLaMA architecture to provide nuanced relevance estimations, thereby enhancing the precision of reranked document lists. Recent works proposed open-source fine-tuned listwise LLM rankers, like \textbf{RankZephyr}~\cite{pradeep2023rankzephyr} and \textbf{RankVicuna}~\cite{pradeep2023rankvicuna}. RankVicuna is built upon Vicuna~\footnote{\url{https://huggingface.co/lmsys/vicuna-7b-v1.5}} model and distilled using RankGPT-3.5~\cite{sun2023chatgpt} as a teacher for some queries from the MSMARCO passage dataset. RankZephyr is an improved version of RankVicuna and further fine-tuned using RankGPT-4~\cite{sun2023chatgpt} only for a smaller set of queries. The initial retrieval is done using BM25 for these rankers. Another notable model, \textbf{LiT5}~\cite{tamber2023scaling}, is based on the T5 architecture and is optimized for listwise ranking tasks. Given that these models exhibit a limited context length, various approaches have been proposed to aggregate sub-lists from an initial ranker into a final ranked list~\cite{qin2023large,parry2024top}.

Despite the advancements introduced by these LLM-based rerankers, integrating adaptive retrieval (AR) techniques with listwise ranking outputs remains a unexplored. Our work addresses these gaps by proposing an adaptive retrieval mechanism specifically designed for LLM rerankers, thereby enhancing both recall and ranking effectiveness.

\section{\sgar: Sliding Window based Adaptive Retrieval}
\label{sec:method}
\subsection{Preliminaries}
Recall from Section~\ref{sec:back-ada} that Graph-based Adaptive Retrieval (\gar{}) and  Query Affinity Modelling based Adaptive Retrieval (\quam{}) augment the set of documents for reranking by selecting the neighbors of top-ranked documents by leveraging a corpus graph (constructed offline) to find neighbors in constant time. To formalize and set the stage for our proposed method, let $R_0$ be the initial retrieval pool, $b$ is the batch size, $c$ is the ranking budget, and $G$ is the corpus graph. 
\gar{} and \quam{} start with taking the top $b$ (batch size for pointwise ranker) documents from the initial retrieval $R_0$ and ranking by a pointwise ranker like MonoT5~\cite{nogueira2020document}. Next, considering the top-ranked documents, the corpus graph $G$ is explored to select $b$ neighboring documents. Subsequent iterations alternate between the initial retrieval and the graph neighborhood until the reranking budget $c$ is met. 

\subsection{The \sgar{} Algorithm}
For a given query $q$, the standard listwise reranking pipeline takes the retrieval results as input. Since the LLM rankers are limited by the context size\footnote{Although recent advances in LLMs enable very large context sizes~\cite{chen2023extending}, we continue to use the strategies proposed in prior work so as to remain comparable.}, an effective sliding window strategy is proposed~\cite{sun2023chatgpt}, based on mainly two hyper-parameters, window size $w$ and step size $b$. The LLM ranker takes $w$ documents (in back-to-first order) and reranks them, and then slides the window by step (of size $b$). The sliding window strategy continues until all documents in the list have been processed and reranked.

\algdef{SE}[DOWHILE]{Do}{doWhile}{\algorithmicdo}[1]{\algorithmicwhile\ #1}%

\begin{figure}[t]
    \vspace{-1cm}
    \begin{algorithm}[H]
    \caption{\sgar: Sliding Window based Graph Adaptive Retrieval}    \label{alg:slidegar}
    \begin{algorithmic}[1]
    \Require Initial ranking $R_0$, window size $w$, step size $b$, budget $c$, corpus graph $G$
    \Ensure Reranked pool $R_1$
    \State $R_1 \gets \emptyset$ \Comment{Reranking results}
    \State $P \gets R_0$ \Comment{Reranking pool}
    \State $F \gets \emptyset$ \Comment{Graph frontier}
    \State $L \gets $ \Call{Select}{top $w$ from $R_0$}  \Comment{Prepare first window} 
    \Do
        \State $B \gets$ \Call{ListwiseRanker}{$L$, subject to $w$ and $b$} \Comment{e.g., RankZephyr}
        \State $R_0 \gets R_0 \setminus B$ \Comment{Discard batch from initial ranking}
        \State $L_1 \gets B[1:b]$ \Comment{Keep top $b$ docs for next window}
        \State $R_1 \gets R_1 \cup  B[b+1:|B|]$ \Comment{Add remaining to results}
        \State $F \gets \Call{Neighbours}{B, G} \setminus (R_1\cup L_1)$ \Comment{Update frontier}
        \State $P \gets \begin{cases} 
        R_0 & \text{if } P = F \\
        F   & \text{if } P = R_0 \\
        \end{cases}$ \Comment{Alternate between initial ranking and frontier}
        \State $L_2 \gets $ \Call{Select}{top $b$ from $P$} 
        \State $L \gets L_1 \cup L_2 $  \Comment{Prepare window for next round}    
    \doWhile{$|R_1| < c-b$} 
    \State $R_1 \gets R_1 \cup L_1$ 
    \end{algorithmic}
    \end{algorithm}
    \vspace{-1.5cm}
\end{figure}

Since \gar{} and \quam{} are based on the pointwise ranker, they are not directly applicable to the listwise rankers. In our work, we propose the \sgar~(Sliding-window-based Graph Adaptive Retrieval) algorithm to use adaptive retrieval for listwise rankers. Algorithm~\ref{alg:slidegar} presents the workings of \sgar{}, also visualized in Figure~\ref{alg:slidegar}. For query $q$, let $R_0$ and $R_1$ be the initial and final pool of ranked documents. The initial pool $R_0$ is the ranking given the retriever, e.g., BM25 and $R_1$ is initialized empty. Let $w$ be the sliding window size and $b$ be the step size. Similar to \gar{}, we consider $P$ as the reranking pool of documents (initialized with $R_0$, $P \gets R_0$), which will be dynamically updated. We consider graph frontier $F$, which will contain graph neighborhood documents. Contrary to the standard sliding window listwise rankers, \sgar{} starts in left-to-right order and takes top $w$ documents (denoted by $L$) from the $R_0$ and ranks them using the listwise ranker (e.g., RankZephyr). Let $B$ be the ranked list (without scores). These $w$ documents are removed from $R_0$ since they are ranked. Next, let $L_1$ be the list of top $b$ documents from $B$ and taken to the next window. We add the remaining documents to $R_1$. Since the ranker does not provide any score, we use the reciprocal of the rank as a pseudo-score to the documents in $B$.

\begin{figure}
    \centering
    \includegraphics[width=0.7\linewidth]{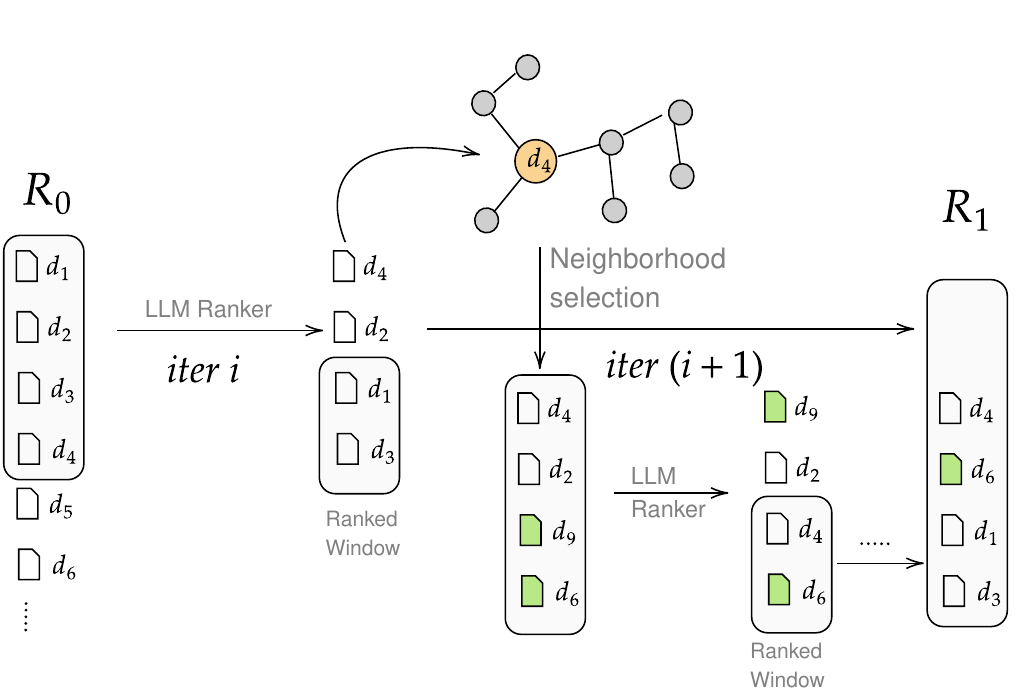}
    
    \caption{The \sgar{} algorithm visualized. LLM Ranker ranks the list of documents (window) and then \sgar{} leverages LLM Ranker feedback and looks for the neighbors of $d_4$ and $d_2$, and carries both documents and their neighbors in the next window. The neighborhood documents are highlighted in green. The remaining documents, $d_1$ and $d_3$, are added to $R_1$.}
    \label{fig:enter-label}
    \vspace{-0.5cm}
\end{figure}

The graph frontier $F$ is now updated with the neighbors (which are not in $R_1$ and $L_1$) of documents from $B$. The neighbors are prioritized by the scores of the source documents in $B$. For the next windows, we follow the conventions proposed by~\cite{macavaney2022adaptive} and alternate between the $R_0$ and the graph frontier $F$. We take the top $b$ documents from $F$ (denoted by $L_2$) and prepare the window for the next ranking round ($L\gets L_1 \cup L_2$). \sgar{} is generalizable for any choice of $w$ and $b$, but for simplicity, we assume that $b=w/2$. We keep alternating between the initial ranking $R_0$ and graph frontier $F$ until $R_1$ has $c-b$ documents and then finally the documents from $L_1$ are appended on the top of $R_1$. Similar to standard listwise LLM ranker, \sgar{} needs  \( \left\lceil \frac{c - w}{b} \right\rceil + 1 \) LLM calls to process $c$ documents with window size $w$ and step size $b$.

\sgar{} has some similarities to existing methods. Compared to \gar{}~\cite{macavaney2022adaptive}, it does not assume scores are independent, making it suitable for listwise re-ranking methods. When compared with TDPart~\cite{parry2024top}, it does not perform partitioning and pulls in documents missing from the initial results.

\section{Experimental Setup}
\label{sec:setup}

Our experiments answer the following research questions about \sgar:
\vspace{-0.8em}
\begin{enumerate}
\item[-] \textbf{RQ1:} What is the impact of adaptive retrieval in the ranking performance of LLM rerankers?
\item[-] \textbf{RQ2:} How sensitive are the LLM based rerankers to the graph depth ($k$)?
\item[-] \textbf{RQ3:} What are the additional computational overheads by \sgar{} in comparison to standard LLM based reranking?
\end{enumerate}
\vspace{-2em}

\subsection{Dataset and Evaluation}
In our experiments, we use the MSMARCO~\cite{bajaj2016ms} passage corpus and evaluate performance on TREC Deep Learning 2019 (DL19) and 2020 (DL20) query sets. The DL19 and DL20 datasets have 43 and 54 queries respectively. To further study the generalization abilities of our method, we use MSMARCO-passage-v2 corpus and test on TREC Deep Learning 2021 (DL21) and 2022 (DL22). Since this corpus contains duplicate documents (same text assigned to different document IDs), we de-duplicate the whole corpus and create BM25~\cite{robertson2009probabilistic} and TCT~\cite{lin2021batch} indexes. We also de-duplicate the qrels from the test sets DL21 and DL22. We release de-duplicate corpus indexes and qrels with our code. The DL21 and DL22 datasets have 53 and 76 queries respectively.

We use nDCG@10 and Recall@c to evaluate different ranking approaches, where $c$ is the reranking budget. We use the PyTerrier~\cite{macdonald2021pyterrier} framework for evaluating runs and follow the standard ranking pipeline notation from PyTerrier, for example, \textbf{BM25>>RankZephyr} represents the ranking pipeline where retrieval is done using BM25 and ranking by RankZephyr model. We re-use the BM25~\cite{robertson2009probabilistic} and TCT~\cite{lin2021batch} based document corpus graphs from \gar{}~\cite{macavaney2022adaptive}, consisting of the $16$ nearest neighbors of each document in the corpus. For MSMARCO-passage-v2, we build both BM25 and TCT-based corpus graphs with $16$ neighbors. We release the corpus graphs with our code. 

\subsection{Retrieval and Ranking Models}
\mpara{Retrievers:} We perform experiments with both sparse and dense retrieval methods. For sparse retrieval, we use BM25~\cite{robertson2009probabilistic} using a PISA~\cite{mallia2019pisa} index. For dense retrieval, we use TCT~\cite{lin2021batch} and retrieve (exhaustively) top $c\in[50,100]$ documents using the TCT-ColBERT-HNP~\cite{lin2021batch} model from \textit{huggingface}\footnote{\url{https://huggingface.co/castorini/tct\_colbert-v2-hnp-msmarco}}.

\mpara{Rankers:} We use different listwise LLM rankers, both zero-shot and fine-tuned. We use zero-shot rankers from \textbf{RankGPT}~\cite{sun2023chatgpt} mainly based on \texttt{gpt-3.5-turbo} and \texttt{gpt-4o} models from OpenAI~\cite{openai2024}. We use Azure API\footnote{\url{https://learn.microsoft.com/en-us/azure/ai-services/openai/overview}} for both \texttt{gpt-3.5-turbo} (gpt-35-turbo in Azure) and \texttt{gpt-4o} models. We also use fine-tuned listwise LLM rankers, mainly \textbf{RankZephyr}~\cite{pradeep2023rankzephyr} and \textbf{RankVicuna}~\cite{pradeep2023rankvicuna}. For all these listwise rankers, we use the implementation from the open-source Python library, called Rerankers~\cite{clavié2024rerankers} which incorporates the RankLLM\footnote{\url{https://github.com/castorini/rank\_llm/tree/main}} (a better variant of RankGPT). These models use the sliding window strategy over a list of documents and generate ranking (also called permutation). It is important to note that these rankers do not provide scores. We use default settings where the sliding window size is $20$ and the step size is $10$. We use the context size of $4096$ for all rankers.

\mpara{Baselines:}
We compare our approach to the baseline (non-adaptive) LLM reranking systems. Furthermore, we compare with a variation of \sgar{} that uses RM3~\cite{abdul2004umass} (denoted by \slidesys{RM3}) as its source of additional relevant documents, rather than the corpus graph. Specifically, after reranking the current window by the LLM, we expand the query using top-$b$ documents and for the subsequent window, retrieve the remaining $b$ documents using the expanded query from the index (mainly BM25). Note that the LLM always uses the original query for reranking; the RM3-expanded query is only used to retrieve additional results. For \slidesys{RM3}, we set both \textit{fb\_docs} and \textit{fb\_terms} to 10, and the \textit{original\_query\_weight} to 0.6.

\mpara{Other Hyperparameters:}
We denote the reranking budget by $c$, window size $w$, step size by $b$, and graph depth (number of neighbors in the corpus graph) by $k$. We choose the reranking budget $c$ of 50 and 100. We set $w$=20, $b$=10 and the depth of the graph $k$=16. For further ablation study, we vary graph depth $k$ in [2,16] (multiples of 2).

\section{Results and Analysis}
We now discuss our results and provide a detailed analysis. We denote our sliding window and graph-based retrieval method by \slidesys{} and the corresponding graph in the subscript, for example, \slidesys{BM25} represents the \slidesys{} with a BM25-based graph. We incorporate \slidesys{} with all ranking methods.

\subsection{Effectiveness}
To answer \textbf{RQ1}, we propose the \slidesys{} algorithm and assess the effectiveness over different reranking budgets, initial retrievers, and rerankers. We report the performance of different reranking approaches in Table~\ref{tab:main_result}. 
We observe that \slidesys{} shows a significant improvement over the standard ranking. Across all rankers, the improved recall remains in the same range for the given reranking budget. However, since we use the window level ranking to look for the neighbors, we find that the better ranker provides slightly better recall. Hence, RankGPT-4o and RankZephyr show slightly better recall improvements than RankGPT-3.5 and RankVicuna. 

\begin{table}[!t]
    \centering
    \caption{The performance comparison of different LLM-based rankers on TREC DL19 and DL20. Significant improvements (paired t-test, $p<0.05$, using Bonferroni correction) with the reranking baseline are marked with * in the superscript. Boldface indicates the strongest result within each group.}
    {\small
    \setlength{\tabcolsep}{2.0pt}
    \begin{tabular}{l|rrrr|rrrr}
        \toprule
        &\multicolumn{4}{c|}{DL19} & \multicolumn{4}{c}{DL20}  \\
        \cmidrule(lr){2-5}\cmidrule(lr){6-9}
        &\multicolumn{2}{c}{$c=50$} &\multicolumn{2}{c|}{$c=100$} & \multicolumn{2}{c}{$c=50$} &\multicolumn{2}{c}{$c=100$} \\
        \cmidrule(lr){2-3}\cmidrule(lr){4-5}\cmidrule(lr){6-7}\cmidrule(lr){8-9}

        pipeline &nDCG &R@c& nDCG& R@c &nDCG&R@c &nDCG& R@c\\
         \toprule

        \rowcolor{gray!50} 
        BM25>>RankGPT-4o & 0.704 & 0.389 &0.745&0.497 &0.688 &0.465 &0.712 & 0.569\\
        w/ \slidesys{RM3} &  0.712&*0.441 &0.729 &*0.538& *0.747& *0.530& *0.760 & *0.660\\        
        w/ \slidesys{BM25} & 0.741 & *0.445&0.742 &*0.543  &0.718 &0.499 &*0.750 &*0.615 \\
         w/ \slidesys{TCT} & \bf0.762&*\bf0.498 &\bf0.771 &*\bf0.601 & *\bf0.779 & *\bf0.578& *\bf0.771 &*\bf0.678 \\
        \midrule 
        \rowcolor{gray!50} 
         BM25>>RankGPT-3.5 & 0.623 &0.389  &0.642  &0.497 & 0.585&0.465 &0.615&0.569 \\
        w/ \slidesys{RM3} & 0.639 & *0.422 &  0.651 &*0.549  &0.596  &*0.512  & 0.599 &*0.600\\       
         w/ \slidesys{BM25} &0.651 &*0.434 & 0.655 & *0.549 &0.604&0.495&0.603&0.597 \\
         w/ \slidesys{TCT} & \bf0.660 &*\bf0.480& \bf0.659   & *\bf0.589  & *\bf0.652& *\bf0.560&\bf0.646&*\bf0.671\\
        \midrule
        \rowcolor{gray!50} 
           BM25>>RankZephyr & 0.670  &0.389 & 0.714 &0.497 &0.684&0.465&0.708 &0.569\\
        w/ \slidesys{RM3} & 0.711 & *0.438  & 0.740 & *0.553 & 0.699 & *0.508 & 0.720& *0.608\\       
         w/ \slidesys{BM25} & 0.705 & *0.430  &0.715 & *0.546  &0.703&0.498 &0.727&*0.612\\
         w/ \slidesys{TCT} & *\bf0.747  &*\bf0.481& \bf0.747 &*\bf0.590 &*\bf0.772&*\bf0.574 & *\bf0.779&*\bf0.681 \\
        \midrule
        \rowcolor{gray!50} 
          BM25>>RankVicuna  & 0.633 & 0.389 &\bf0.663&0.497 &0.646&0.465&0.663&0.569 \\
        w/ \slidesys{RM3} & 0.645 &0.418   & 0.658& 0.528 & 0.653&*0.515&  0.663 & 0.593 \\        
         w/ \slidesys{BM25} & 0.650  &*0.427 & 0.650&*0.543  & 0.639 &0.485 &0.654&0.596\\
         w/ \slidesys{TCT} & \bf0.655 &*\bf0.478 & 0.661 &*\bf0.571  & \bf0.661&*\bf0.570 &\bf0.677&*\bf0.672\\
        \midrule
        \midrule
        \rowcolor{gray!50} 
          TCT>>RankGPT-4o & 0.792 &0.506 & 0.775 & 0.610 & 0.779&\bf0.652 & \bf0.787& 0.713  \\
            w/  \slidesys{BM25} & 0.809 &0.541 & \bf0.793& 0.645  & \bf0.783 & 0.638 & 0.771 & 0.723 \\
           w/ \slidesys{TCT} & \bf0.816&*\bf0.561  & 0.771 &  *\bf0.666 &0.777 &0.643  & 0.755 & *\bf0.733 \\
        \midrule
        \rowcolor{gray!50} 
          TCT>>RankGPT-3.5  &0.755  &0.506 &0.753&0.610  &\bf0.736&\bf0.652 &\bf0.738&0.713 \\
            w/  \slidesys{BM25} &\bf0.765  & 0.525&0.747&0.641 &0.721&0.641 &0.702&*\bf0.737\\
           w/  \slidesys{TCT}  &0.758&*\bf0.539& \bf0.765 &*\bf0.647 &0.697&0.645&0.680&*0.729\\
        \midrule
        \rowcolor{gray!50} 
        TCT>>RankZephyr  &0.771&0.506  &0.763 &0.610 &0.772&\bf0.652&\bf0.772&0.713\\
        w/  \slidesys{BM25}   &0.787 &0.534  &0.788&\bf0.655  &0.768&0.638&0.766&\bf0.734\\
             w/  \slidesys{TCT}  &\bf0.800&*\bf0.552 & \bf0.789&*0.653 &\bf0.775&0.645&0.770&*0.728\\
         \midrule                 
        \rowcolor{gray!50} %
         TCT>>RankVicuna  &0.746& 0.506 &0.746&0.610   &0.729&\bf0.652&0.719&0.713 \\
           w/  \slidesys{BM25}  &0.756&0.520 & 0.741&0.636  &\bf0.736&0.641&\bf0.736&0.732  \\
           w/  \slidesys{TCT}  &\bf0.763&*\bf0.542 & \bf0.754&*\bf0.647   &0.733&0.649& 0.718&*\bf0.730 \\
     \bottomrule    
    \end{tabular}
    }
    \label{tab:main_result}
\end{table}

The most significant improvements are at the ranking budget of 50 and when the sparse retriever is used. On DL19, Recall@50 improves between 22\% and 28\% across rankers, where BM25>>RankGPT-4o with \slidesys{TCT} shows the greatest improvements. Additionally, the fine-tuned RankZephyr improves nDCG@10 the most for the BM25>>RankZephyr (from 0.670 to 0.747) with \slidesys{TCT}. Recall@100 improves the highest (up to 20\%) for RankGPT-4o and the lowest (up to 14\%) for RankVicuna.  We observe similar trends for the DL20 dataset. Our adaptive RM3 baseline, \slidesys{RM3}, also improves recall (significant in 13 of 16 comparisons), but is not able to successfully re-rank these documents to yield an improved nDCG@10. This observation suggests that the corpus graph exploration approach is preferable to that of traditional PRF variants in terms of both efficiency and effectiveness.
We also find that \slidesys{TCT} shows better performance in comparison to \slidesys{BM25} and \slidesys{RM3}, since the dense retriever-based corpus graph is capable of providing complementary signals to the lexically retrieved documents.

When the dense retriever, TCT, is used for the initial ranking, we continue to see improvements, though the relative differences are lower due to the stronger initial result set. Both RankGPT-4o and RankZephyr improve recall@50 between 9\%-10\% and RankVicuna and RankGPT-3.5 between 6\%-7\% on the DL19. Surprisingly, the nDCG@10 improved from 0.792 to 0.816 for the RankGPT-4o and from 0.771 to 0.800 for RankZephyr with \slidesys{TCT}, achieving the highest reported effectiveness that we are aware of on this benchmark. This demonstrates the ranking capabilities of the LLM-based rerankers and the adaptive retrieval helps in finding missing relevant documents. We see statistically significant improvements in Recall@100 for all rankers with \slidesys{TCT}. However, the Recall@50 does not improve for the DL20 dataset. We suspect that the reason for this observation is twofold. First, the absolute value of the recall in DL20 with TCT is higher than that in DL19, giving less potential room for improvement. This problem is also exacerbated by a lower average number of relevant documents per query in DL20 than DL19 (30.9 vs 58.1). Second, the LLM rerankers may be up against their maximum possible effectiveness on the collection; some relevant documents may be incorrectly ranked poorly by these models, thereby never allowing them to be used as a feedback signal to find additional ones. This observation is challenging to validate, however, because it would require exhaustively scoring the MSMARCO corpus to see where the remaining relevant documents are ranked---which would be very expensive.

Overall, these results indicate that adaptive retrieval techniques, particularly \slidesys{TCT}, substantially enhance the performance of listwise LLM rankers.

\subsection{Generalization on Different Corpus }
In Table~\ref{tab:results-msv2}, we report the results on the MSMARCO-passage-v2 corpus using DL21 and DL22 as test sets. We use RankGPT-3.5 and RankZephyr as rankers. Unsurprisingly, \sgar{} also shows significant improvements on the new corpus. Specifically, on DL21, \slidesys{TCT} improves Recall@50 from 0.248 to 0.356 (43\% improvement) with BM25>>RankGPT-3.5 and from 0.248 to 0.365 (47\% improvement) with BM25>>RankZephyr. Similar improvements are observed when the budget is set to 100. Our method also shows robust performance on the DL22 dataset, where the initial retrieval effectiveness is considerably lower (Recall@50 is 0.119 and Recall@100 is 0.171). 
In particular, \slidesys{TCT} improves Recall@50 by 28\% with BM25>>RankGPT-3.5 and by 47\% with BM25>>RankZephyr. Similar trends are observed when the budget increases to 100. Additionally, recall improvements are evident when the dense (TCT) retriever is used at the initial stage. Overall, we observe better ranking and recall improvements when the fine-tuned listwise ranker, RankZephyr, is utilized.

\begin{table}[h]
    \centering
    \caption{The performance comparison of different LLM-based rankers on TREC DL21 and DL22. Significant improvements (paired t-test, $p<0.05$, using Bonferroni correction) with the reranking baseline are marked with * in the superscript. Boldface indicates the strongest result within each group.}
    {\small
    \setlength{\tabcolsep}{2.0pt}
    \begin{tabular}{l|rrrr|rrrr}
        \toprule
        &\multicolumn{4}{c|}{DL21} & \multicolumn{4}{c}{DL22}  \\
        \cmidrule(lr){2-5}\cmidrule(lr){6-9}
        &\multicolumn{2}{c}{$c=50$} &\multicolumn{2}{c|}{$c=100$} & \multicolumn{2}{c}{$c=50$} &\multicolumn{2}{c}{$c=100$} \\
        \cmidrule(lr){2-3}\cmidrule(lr){4-5}\cmidrule(lr){6-7}\cmidrule(lr){8-9}

        pipeline &nDCG &R@c& nDCG& R@c &nDCG&R@c &nDCG& R@c\\
         \toprule
        \rowcolor{gray!50} 
        BM25>>RankGPT-3.5 & 0.579 &0.248  &\bf0.591 &0.337 & 0.381 &0.119  &\bf0.415 &0.171 \\
        w/ \slidesys{RM3} &0.572 & *0.273 &0.579 &0.361 &0.379 &0.123 &0.395  &0.180 \\ 
        w/ \slidesys{BM25} & 0.580 &*0.328  &0.574  &*0.414 &0.368 &0.128 &0.377  & 0.184\\        
        w/ \slidesys{TCT} & \bf0.594 &*\bf0.356 &0.586 &*\bf0.458  &\bf0.386 &*\bf0.153 &0.396  &*\bf0.225 \\   
        \midrule
        \rowcolor{gray!50} 
        BM25>>RankZephyr & 0.638 &0.248  &0.663 &0.337 & 0.439 &0.119  & 0.486 &0.171 \\
        w/ \slidesys{RM3} & 0.652 &*0.282  & 0.678&*0.371 &0.458 &0.132 &0.492  &*0.192 \\ 
        w/ \slidesys{BM25} & 0.671 &*0.325  &0.685  &*0.429 &0.418 &0.131 &0.459 &*0.192 \\        
        w/ \slidesys{TCT} &*\bf0.692&*\bf0.365 & \bf0.699 &*\bf0.481 &*\bf0.490&*\bf0.175 &  \bf0.518 &*\bf0.266 \\   
        \midrule
        \midrule
        \rowcolor{gray!50} 
        TCT>>RankGPT-3.5 &\bf0.695 &0.406  & \bf0.695&0.527 & \bf0.585 &0.274  &\bf0.589&0.356 \\
        w/ \slidesys{BM25} &0.692 &0.427  &0.693 &*0.565 & 0.556&0.250 &0.544&0.357 \\      
        w/ \slidesys{TCT} &0.677 &*\bf0.439  & 0.658 &*\bf0.584 & 0.552&\bf0.284 &0.532 &\bf0.376 \\   
        \midrule
        \rowcolor{gray!50} 
        TCT>>RankZephyr &0.733 &0.406  & 0.734& 0.527& 0.652 &0.274  &\bf0.663  & 0.356\\
        w/ \slidesys{BM25} &\bf0.759 &0.424  &\bf0.760 &*0.569 &0.657 &0.256 &0.662  &0.362 \\       
        w/ \slidesys{TCT} & 0.745 &*\bf0.445  & 0.753 &*\bf0.576 &\bf0.661  & *\bf0.297 &0.656 &*\bf0.388 \\   
     \bottomrule    
    \end{tabular}
    \vspace{-0.3cm}
    }
    \label{tab:results-msv2}
\end{table}

\subsection{Effect of graph depth}
\begin{figure}[h]
        \begin{subfigure}[b]{0.33\textwidth}
        \begin{tikzpicture}
		\begin{axis}[
			width  = 1.2\linewidth,
			height = 1.2\linewidth,
            mark options={solid}, %
			major x tick style = transparent,
			grid = major,
		    grid style = {dashed, gray!20},
			xlabel = {graph neighbours $k$},
			ylabel = {},
			title={nDCG@10},
            title style={yshift=-1.5ex}, %
            symbolic x coords={2,4,6,8,10,12,14,16},
            xtick={2,4,6,8,10,12,14,16},
            xtick distance=20,
            yticklabel style = {font=\tiny,xshift=0.5ex},
            xticklabel style = {font=\tiny,yshift=0.5ex},
            enlarge x limits=0.05,
            xlabel near ticks,
            ylabel near ticks,
            every axis x label/.style={at={(0.5, -0.08)},anchor=near ticklabel},
            every axis y label/.style={at={(-0.17, 0.5)},rotate=90,anchor=near ticklabel},
			]
   
			\addplot [color=red!60!white, mark=*, mark size=1.0pt,line width = 1.0pt] table [x index=0, y index=1, col sep = comma] {plots/ndcg10c50_gbm25_bm25_dl19_sgar.txt};
			
			\addplot [color=red!60!white, mark=*, style=densely dashed, mark size=1.0pt, line width = 1.0pt] table [x index=0, y index=2, col sep = comma] {plots/ndcg10c50_gbm25_bm25_dl19_sgar.txt};

     		\addplot [color=red!60!white, style=densely dotted, mark=*,mark size=1.0pt, line width = 1.0pt] table [x index=0, y index=3, col sep = comma] {plots/ndcg10c50_gbm25_bm25_dl19_sgar.txt};
        
			\addplot [color=brown,  mark=square*, mark size=1.0pt, line width = 1.0pt] table [x index=0, y index=4, col sep = comma] {plots/ndcg10c50_gbm25_bm25_dl19_sgar.txt};

            \addplot [color=brown, style=densely dashed, mark=square*,mark size=1.0pt, line width = 1.0pt] table [x index=0, y index=5, col sep = comma] {plots/ndcg10c50_gbm25_bm25_dl19_sgar.txt};   

            \addplot [color=brown, style=densely dotted, mark=square*,mark size=1.0pt, line width = 1.0pt] table [x index=0, y index=6, col sep = comma] {plots/ndcg10c50_gbm25_bm25_dl19_sgar.txt};   
            
			\addplot [color=green!60!black,  mark=triangle*, mark size=1.0pt, line width = 1.0pt] table [x index=0, y index=7, col sep = comma] {plots/ndcg10c50_gbm25_bm25_dl19_sgar.txt};

   			\addplot [color=green!60!black,style=densely dashed, mark=triangle*,mark size=1.0pt, line width = 1.0pt] table [x index=0, y index=8, col sep = comma] {plots/ndcg10c50_gbm25_bm25_dl19_sgar.txt};
      
   			\addplot [color=green!60!black, style=densely dotted, mark=triangle*,mark size=1.0pt, line width = 1.0pt] table [x index=0, y index=9, col sep = comma] {plots/ndcg10c50_gbm25_bm25_dl19_sgar.txt};

		\end{axis}
	\end{tikzpicture}%
        \end{subfigure}
	    \begin{subfigure}[b]{0.33\textwidth}
        \begin{tikzpicture}
		\begin{axis}[
            mark options={solid}, %
			width  = 1.2\linewidth,
			height = 1.2\linewidth,
			major x tick style = transparent,
			grid = major,
		    grid style = {dashed, gray!20},
			xlabel = {graph neighbours $k$},
			ylabel = {},            
			title={nDCG@50},
            title style={yshift=-1.5ex}, %
            legend columns=3,
            legend entries={RankVicuna;,RankVicuna w/ \slidesys{{\scalebox{0.7}{BM25}}};,RankVicuna w/ \slidesys{{\scalebox{0.7}{TCT}}};,RankZephyr;, RankZephyr w/\slidesys{{\scalebox{0.7}{BM25}}};,RankZephyr w/\slidesys{{\scalebox{0.7}{TCT}}};,RankGPT-3.5;,RankGPT-3.5  w/ \slidesys{{\scalebox{0.7}{BM25}}};,RankGPT-3.5  w/ \slidesys{{\scalebox{0.7}{TCT}}} },
            legend style={font=\fontsize{6.0}{10}\selectfont, mark options={solid}},
            legend to name=sgar_dl19_named,
            symbolic x coords={2,4,6,8,10,12,14,16},
            xtick={2,4,6,8,10,12,14,16},
            yticklabel style = {font=\tiny,xshift=0.5ex},
            xticklabel style = {font=\tiny,yshift=0.5ex},
            xlabel near ticks,
            ylabel near ticks,
            every axis x label/.style={at={(0.5, -0.08)},anchor=near ticklabel},
            every axis y label/.style={at={(-0.17, 0.5)},rotate=90,anchor=near ticklabel},
            enlarge x limits=0.05,
			]

			\addplot [color=red!60!white, mark=*, line width = 1.0pt,mark size=1.0pt] table [x index=0, y index=1, col sep = comma] {plots/ndcgc50_gbm25_bm25_dl19_sgar.txt};
			
			\addplot [color=red!60!white, mark=*, style=densely dashed, mark size=1.0pt,line width = 1.0pt] table [x index=0, y index=2, col sep = comma] {plots/ndcgc50_gbm25_bm25_dl19_sgar.txt};

			\addplot [color=red!60!white, mark=*,style=densely dotted, mark size=1.0pt,line width = 1.0pt] table [x index=0, y index=3, col sep = comma] {plots/ndcgc50_gbm25_bm25_dl19_sgar.txt};
   
   			\addplot [color=brown,  mark=square*,mark size=1.0pt, line width = 1.0pt] table [x index=0, y index=4, col sep = comma] {plots/ndcgc50_gbm25_bm25_dl19_sgar.txt};
			
            \addplot [color=brown, mark=square*, style=densely dashed, mark size=1.0pt, line width = 1.0pt] table [x index=0, y index=5, col sep = comma] {plots/ndcgc50_gbm25_bm25_dl19_sgar.txt};

            \addplot [color=brown,style=densely dotted, mark=square*, mark size=1.0pt, line width = 1.0pt] table [x index=0, y index=6, col sep = comma] {plots/ndcgc50_gbm25_bm25_dl19_sgar.txt};
            
   			\addplot [color=green!60!black, mark=triangle*,mark size=1.0pt, line width = 1.0pt] table [x index=0, y index=7, col sep = comma] {plots/ndcgc50_gbm25_bm25_dl19_sgar.txt};

   			\addplot [color=green!60!black,  style=densely dashed, mark=triangle*, mark size=1.0pt, line width = 1.0pt] table [x index=0, y index=8, col sep = comma] {plots/ndcgc50_gbm25_bm25_dl19_sgar.txt};

   			\addplot [color=green!60!black, style=densely dotted, mark=triangle*, mark size=1.0pt, line width = 1.0pt] table [x index=0, y index=9, col sep = comma] {plots/ndcgc50_gbm25_bm25_dl19_sgar.txt};
      
		\end{axis}
	\end{tikzpicture}
        \end{subfigure}
          \hspace{-0.01\columnwidth}
        \begin{subfigure}[b]{0.33\textwidth}
            \begin{tikzpicture}
		\begin{axis}[
			width  = 1.2\linewidth,
			height = 1.2\textwidth,
            mark options={solid}, %
			major x tick style = transparent,
			grid = major,
		    grid style = {dashed, gray!20},
			xlabel = {graph neighbours $k$},
			ylabel = {},            
			title={Recall@50},
            title style={yshift=-1.5ex}, %
            symbolic x coords={2,4,6,8,10,12,14,16},
            xtick={2,4,6,8,10,12,14,16},
            xtick distance=20,
            enlarge x limits=0.05,
            yticklabel style = {font=\tiny,xshift=0.5ex},
            xticklabel style = {font=\tiny,yshift=0.5ex},
            xlabel near ticks,
            ylabel near ticks,
            every axis x label/.style={at={(0.5, -0.08)},anchor=near ticklabel},
            every axis y label/.style={at={(-0.17, 0.5)},rotate=90,anchor=near ticklabel},
			]
			
			\addplot [color=red!60!white,mark=*,mark size=1.0pt, line width = 1.0pt] table [x index=0, y index=1, col sep = comma] {plots/recallc50_gbm25_bm25_dl19_sgar.txt};

            \addplot [color=red!60!white, mark=*, style=densely dashed, mark size=1.0pt, line width = 1.0pt] table [x index=0, y index=2, col sep = comma] {plots/recallc50_gbm25_bm25_dl19_sgar.txt};

            \addplot [color=red!60!white, style=densely dotted, mark=*, mark size=1.0pt, line width = 1.0pt] table [x index=0, y index=3, col sep = comma] {plots/recallc50_gbm25_bm25_dl19_sgar.txt};
            
            \addplot [color=brown, mark=square*,mark size=1.0pt, line width = 1.0pt] table [x index=0, y index=4, col sep = comma] {plots/recallc50_gbm25_bm25_dl19_sgar.txt};
            
            \addplot [color=brown, mark=square*, style=densely dashed, mark size=1.0pt, line width = 1.0pt] table [x index=0, y index=5, col sep = comma] {plots/recallc50_gbm25_bm25_dl19_sgar.txt};

            \addplot [color=brown, style=densely dotted, mark=square*,mark size=1.0pt, line width = 1.0pt] table [x index=0, y index=6, col sep = comma] {plots/recallc50_gbm25_bm25_dl19_sgar.txt};
            
            \addplot [color=green!60!black, mark=triangle*,mark size=1.0pt, line width = 1.0pt] table [x index=0, y index=7, col sep = comma] {plots/recallc50_gbm25_bm25_dl19_sgar.txt};

   			\addplot [color=green!60!black, style=densely dashed, mark=triangle*, mark size=1.0pt, line width = 1.0pt] table [x index=0, y index=8, col sep = comma] {plots/recallc50_gbm25_bm25_dl19_sgar.txt};

     		\addplot [color=green!60!black, style=densely dotted, mark=triangle*, mark size=1.0pt, line width = 1.0pt] table [x index=0, y index=9, col sep = comma] {plots/recallc50_gbm25_bm25_dl19_sgar.txt};
      
		\end{axis}
	\end{tikzpicture}
        \end{subfigure}                 
                
\begin{minipage}{\textwidth}
\begin{center}
\ref{sgar_dl19_named}
\end{center}
\end{minipage}   
 \normalsize
    \vspace{-0.8em}
    \caption{Effect of \sgar{} on different ranking pipelines on \textbf{TREC DL19} dataset when the number of graph neighbors $k$ varies and window size is 20 and step size 10. } 
    \label{fig:ablation-sgar-dl19}
\end{figure}
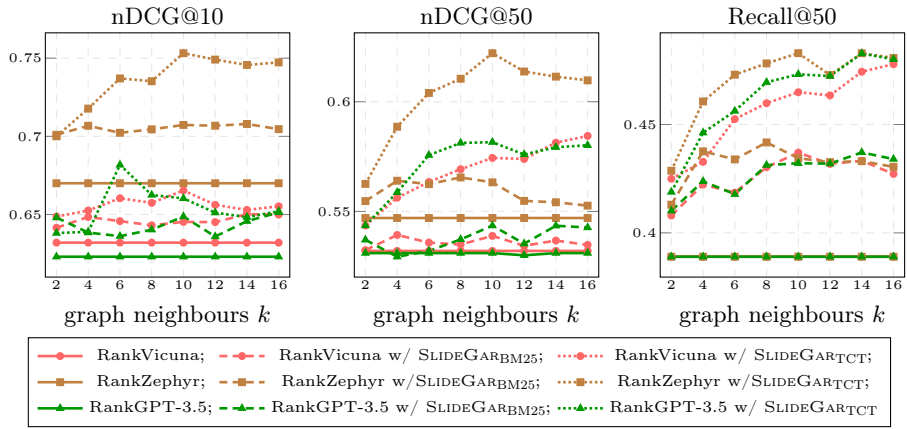

A key hyperparameter of \sgar{} is the graph depth $k$, which we now investigate to answer \textbf{RQ2}.
The adaptive retrieval is mainly based on the number of neighbors, $k$, in the corpus graph. In this section, we study the behavior of different ranking approaches when graph depth $k$ varies. We do not consider RankGPT-4o for this ablation study due to its high cost. Towards this, we vary the graph depth $k\in[2,16]$ (by multiples of 2) and fix the reranking budget $c$ to 50. We report the performance across different metrics in Figure~\ref{fig:ablation-sgar-dl19}. We observe that the recall@50 improves for \slidesys{} as we go deeper into the graph neighborhood. Though, the BM25 variant, \slidesys{BM25} seems to be robust to the number of neighbors. A similar trend can be seen in the nDCG metrics. On the other hand, the \slidesys{TCT} is able to find further relevant documents as we go deeper in the graph. In particular, recall@50 improves for both RankVicuna and RankGPT-3.5, and a similar effect can be seen in nDCG@50 even though the nDCG@10 does not vary. Unsurprisingly, RankZephyr shows good improvements as $k$ starts from 2 to 10 and then becomes stable.  Since, in the alternate iteration, \slidesys{} selects $b$ (step size, which we set to 10) documents from the neighborhood, \slidesys{} might end up taking all $b$ documents from a single source document (the top-ranked document) from the current window.

\subsection{Computational Overhead}
We re-iterate that \slidesys{} alternate between initial retrieved documents and neighborhood documents after each window is processed and has the same number of reranking operations (number of LLM calls depending upon window size $w$ and step size $b$) as standard listwise ranking. The cost of ranking all windows dominates the total cost of reranking the pipeline (including the cost of preparing windows). To answer \textbf{RQ3}, we are interested to see what the additional latency cost our adaptive sliding window approach, \sgar{}, adds to the total LLM reranking latency. Toward this, we conduct an experiment on the DL19 dataset with the reranking budget $c$ of $50$ and $100$. We use BM25 as initial retrieval and \slidesys{BM25} variant of our adaptive method. In Table~\ref{tab:latency}, we report the mean latency (in ms) per query for reranking and \slidesys{} component during reranking. For this experiment, we only use RankZephyr and RankVicuna rankers and do not consider the API-based rankers. We report the mean and variance over $5$ consecutive runs. We conduct our experiment on the NVIDIA V100 GPU with 32G RAM. We observe that RankVicuna and RankZephyr take around 13797 ms and 14884 ms, respectively, to process 50 documents per query. On the other hand, the \slidesys{} component adds only around 2.6ms which is just 0.02\% of the latency cost of LLM ranker. Similarly, when reranking 100 documents, \slidesys{} adds negligible latency.  Overall, our adaptive retrieval approach adds a minimal latency cost while providing better recall.

\begin{table}
    \centering
    \caption{Mean Latency (ms/query) of LLM Rerankers at $c=50$ and $c=100$.}
    \setlength{\tabcolsep}{3.0pt}
    \begin{tabular}{ccccc}
        \toprule
        & \multicolumn{2}{c}{RankVicuna}  & \multicolumn{2}{c}{RankZephyr} \\
        \cmidrule(lr){2-3}\cmidrule(lr){4-5}
        budget &  Ranking &\slidesys{} &Ranking  & \slidesys{}\\
         \midrule
         50 & $13797\pm2.57$ &$2.63\pm0.01$ &$14884\pm1.90$ & $2.67\pm0.01$\\
         100 & $30923\pm0.89$ &$6.73\pm0.01$  &$33474\pm7.04$ &$6.97\pm0.01$ \\
         \bottomrule
    \end{tabular}
        \vspace{-1cm}
    \label{tab:latency}
\end{table}
\section{Conclusion}

We augment existing adaptive ranking algorithms to work with listwise LLM reranking models. 
We find that our proposed method, \sgar{}, is able to successfully overcome the \textit{bounded recall problem} from first-stage retrievers by successfully leveraging feedback signals from an LLM. Also, the computational overhead of applying \sgar{} is minimal compared to a typical LLM reranking pipeline. In our opinion this work enables the broader adoption of LLM reranking, such as in cases where the first stage is unsuccessful or systems are limited by legacy first-stage (lexical) keyword-based retrieval systems.

\bibliographystyle{splncs04}
\bibliography{references}

\end{document}